\documentclass[12pt]{iopart}

\newcommand*{\doi}[1]{\href{http://dx.doi.org/#1}{doi: #1}}
\usepackage{iopams}  
\usepackage{graphicx}
\usepackage{hyperref}
\begin{document}

\title[A Ball Pool model for the Higgs]{A Ball Pool Model to illustrate the Higgs physics to the public}

\author{Giovanni Organtini}

\address{"Sapienza", Universit\`a di Roma \& INFN-Sez. di Roma, Roma, P.le A. Moro 2 I-00185}
\ead{giovanni.organtini@roma1.infn.it}
\begin{abstract}
A simple model is presented to explain the Higgs boson physics to the grand public. The model consists of a children ball pool representing a Universe filled with a certain amount of the Higgs field. The model is suitable for usage as a hands-on tool in scientific exhibits and provides a clear explanation of almost all the aspects of the physics of the Higgs field interaction with other particles. 
\end{abstract}

\pacs{01.20.+x, 01.50.My, 01.40.-d, 14.80.Bn}
\submitto{Physics Education}
\maketitle

\section{Introduction}\label{sec:introduction}
The Higgs boson was discovered in 2012 by the ATLAS and CMS Collaborations at LHC~\cite{ATLAS}~\cite{CMS}. The physics of the Higgs boson interactions with other particles and with itself has been elucidated by few qualitative models, suitable to illustrate it to undergraduate students and to the grand public.  A very popular model is the one formulated by David Miller of University College London (for which he won a champagne bottle awarded by the UK Science Minister William Waldegrave  intended for those who would be able to explain the Higgs mechanism to the people)~\cite{miller}.

Since then a number of analogies were proposed to help people understand the Higgs mechanism, most of which relies on the analogy of the Higgs field with a viscous medium in which particles move slower than in air. Such an analogy provides an effective way to illustrate how the interaction of a field with a massless particle gives rise to the mass of the latter: the mass appears as a reduction of the particle speed in {\em vacuum}, where the {\em vacuum} is a state with the minimum possible energy.

These analogies do not face the problem of explaining the nature of the vacuum state that does not correspond to an empty space, but to a space filled with a certain amount of Higgs field for which the Higgs auto-interaction potential attains its minimum. They also fail in explaining what it means to {\em observe} a Higgs boson.

In our recent works we provided two alternative models addressing these problems. One is a video (in Italian)~\cite{organtinivideo} illustrating the interaction of massless particles, represented by steel balls, with the Higgs field represented by magnets. In this video a massless particle interacts with the field that {\em limits} its ability to move in a given region of space, depending on the nature of the particle (i.e., on the coupling constant between the field and the particle). Another resource is a paper~\cite{unveiling} illustrating the Higgs mechanism using a completely classical formalism, suitable to be understood by most of teachers (even those who does not know about quantum mechanics) and at least part of the students.

None of the models cited before is suitable for the realisation of a {\em hands on} installation, as it may be required by scientific museums or institutions. In this paper we propose a model that is very attractive because it permits to realise an installation in which people can immerse themselves to directly experiment the interaction with a Higgs boson. In the following sections we illustrate the installation and its usage as a pedagogical tool to address all the aspects of the interaction of the Higgs field with massless particles, namely:

\begin{enumerate}
\item contrary to familiar fields, when the Higgs potential $V$ attains its minimum value $V_0 = V\left(\phi_0\right)$ in a volume there is, in fact, a certain amount of field $\phi_0 \ne 0$ that, however, is unobservable; 
\item if the amount of the Higgs field $\phi=\phi_0+\eta$ inside this volume is reduced or increased, the potential changes and the Higgs field becomes observable (this is an optional feature);
\item a massless particle in such a field moves with a speed lower than the speed of light, just because it interacts with the Higgs field;
\item the interaction of matter fields with the Higgs field may led to the observation of a Higgs boson.
\end{enumerate}
Section~\ref{sec:setup} describes the setup of the installation;  Section~\ref{sec:method} shows how to use the installation to illustrate the properties of the Higgs boson.

\section{The setup}\label{sec:setup}
\begin{figure}
\begin{center}
\includegraphics[width=0.45\textwidth]{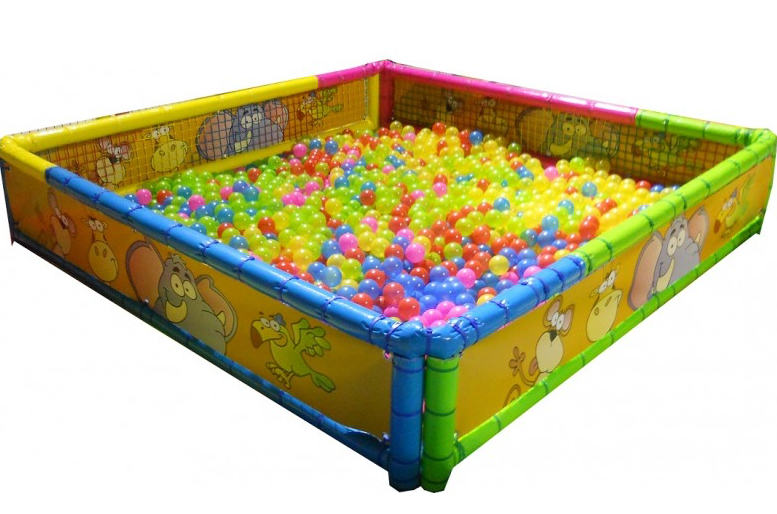}
\caption{\label{fig:ballpool}A children ball pool is a good model of a Universe filled by a certain amount of Higgs field (image is a courtesy of Tutto Gonfiabili -- Play Wily).}
\end{center}
\end{figure}

A very simple setup can be realised with a children ball pool, such as in Fig.~\ref{fig:ballpool}. The walls of the pool must be opaque and the pool must be partially filled. It should stay at a height such that people looking from outside cannot see its content. Optionally, the pool must sit on a scale through which its weight can be measured. The sensitivity of the scale must be good enough to make it possible to appreciate a variation in weight due to the removal of few balls and must indicate zero in normal conditions.

The pool represents the Universe, while the balls represent the Higgs field permeating the Universe. The balls have a nice feature in this context: while acting almost as a fluid when treated as a whole, they are discrete objects. Such a feature provides a very good support to explain the dual nature of quantum fields that behave, at the same time, as waves and particles.

The setup is easy to be realised, cost effective and clean (no liquids required).

\section{The illustration of the Higgs physics}\label{sec:method}
In this section we illustrate the various aspects of the Higgs physics using our model of the Universe, represented by the pool, and of the Higgs field, represented by the balls.

\subsection{The Higgs field at its minimum possible energy}
The first feature of the Higgs field, listed in Section~\ref{sec:introduction}, namely the fact that the Higgs field is unobservable when it attains its minimum energy, is shown looking at the pool from outside. The balls cannot be seen and a {\em scientist} can only conclude that the Universe, represented by the pool, is empty. However, we do know that such a Universe is not empty: there is, actually, some {\em field} inside. The only way to try to detect this field is to interact with it.

\subsection{Changing the amount of the fields changes the potential}
In the Higgs mechanism, the field $\phi$ interacts with itself with a potential reaching its minimum for a value $\phi_0\ne 0$. That means that the state of the lowest possible energy consists in a state in which some Higgs field is present, contrary to what happens in the case of the more familiar electromagnetic field, for which the minimum possible energy is attained when both the electric field $E$ and the magnetic field $B$ are zero.

A consequence of this behaviour is the following. In classical physics, if a given volume is {\em empty} its energy attains its lowest possible value. Such a state is called the {\em vacuum} state meaning that, to increase its energy, something must be {\em added} to the volume. In the case of the Higgs field, the lowest possible value of the energy in a volume is obtained when the volume contains a field $\phi=\phi_0$. Adding some Higgs field to such a volume increases its energy, as for the electromagnetic field. However, in this case, the energy inside the volume increases even when the Higgs field is removed from the volume. For this reason the vacuum state, considered as the state with the lowest possible energy, no longer coincide with an empty state.

In order to detect the presence of some field or particle in a volume we must interact with that field or particle. As a result the energy of the field or particle changes. In the case of the Higgs field, the only way to detect its presence consists in interacting with it causing an increase of the energy in the volume in which is contained. For this reason it is impossible to {\em see} the field $\phi_0$: each time one tries to interact with it, the energy in the volume increases and that implies an increase of the field $\phi>\phi_0$.

Removing or adding some extra field to our setup means removing or adding balls from it. This changes the weight of the pool and the effect can be read on the scale. When the scale is not at zero the system is not in equilibrium. Correspondingly one can {\em see} the removed balls or the added ones from outside, concluding that there is something else in the region out of the vacuum. The only way to return the system to the stable condition is to bring it back to the original condition of zero weight. Here, the weight of the system plays the role of the Higgs potential. When such a potential is null the system is at its equilibrium and this happens when there is a specific amount of field inside the volume, that cannot however be detected. The detection of some extra Higgs field implies a non vanishing potential.

This analogy is only partial, in fact. While the Higgs potential always increases, irrespective of the sign of $\eta$, the weight of the system either increases or decreases. However, the scale can be configured such that it only shows the absolute value of the difference in weight and this recover the analogy.

Moreover, the Higgs field would spontaneously recover the equilibrium, if the field content changes, while in this case this does not happen. Manifestly, the proposed setup is only a metaphor of the Higgs field and thus cannot be fully consistent with it.

As outlined above, this part of the exhibit can optionally be ignored, being also rather difficult to control, especially when the device is not supervised by experts. This feature makes the setup very flexible and suitable for different environmental conditions.

\subsection{The interaction with matter particles}
The mass of the fermions in the Standard Model (SM) arise from the Yukawa interaction of otherwise massless particles with the Higgs field at its minimum value~\cite{PDG}. In other words, the mass of a particle ceases to be a {\em property} of that particle, like its electric charge, but arises from dynamical effects consisting in the interaction of the Higgs field with the particle itself that, in this model, would be massless if not interacting with such a field.

Explaining how an interaction can lead to the appearance of a mass for a particle has been the goal of most of the metaphors developed so far (see, e.g.~\cite{miller}).

In our analogy, a massless particle is represented by a person moving in the space around our Universe consisting of the pool. Those people can move at their own speed without limitations. If the same person walks inside the Universe filled with the Higgs field, its speed is reduced, as if it acquires inertial mass. On the other hand, inertial mass is a measure of how difficult is to make a particle to accelerate. The reduction of the particle speed depends on the fact that now the {\em particle} interacts with the {\em field}. 

Different persons may interact differently with the balls, as different particles do with the Higgs field, leading to particles with different mass. A muon differs from an electron just because its interaction with the Higgs field is stronger than that of the electron. A tall person may move faster than a short boy in the ball filled Universe, because the degree of interaction of the latter is higher.

Experimenting the effect by themselves makes it extremely clear one of the most difficult concepts to understand in Higgs physics: how the interaction with a field can lead to a mass for an otherwise massless particle.

In summary, people observing a person (representing a massless particle) walking inside the pool can only measure the speed of it (the particle) to be lower than that would be outside the {\em Universe}. 

The conclusion is that {\em particles} that move at the same speed in an empty Universe, move with different speeds in a Universe filled with some Higgs field at its lowest energy state, that however appears as empty since such a field is not observable. The speed of each particle depends on the coupling between the Higgs field and the particle: the stronger the coupling, the lower the speed, hence the higher the inertial mass.

\subsection{Observing a Higgs boson}
When a person moves inside the pool, people can only see part of its body. They do not see the {\em field}: it is still not possible to observe any Higgs boson in these conditions. 

In order to observe a Higgs boson in an experiment what we need to do is to provide enough energy to the field such that it can {\em materialise} into a particle that eventually will disappear decaying into a pair of other particles. As a matter of fact, Higgs bosons can be produced at LHC because the energy of the collision between protons turns into the mass of the Higgs boson plus other particles. In other words, one need to transfer enough energy to the vacuum to produce real observable Higgs bosons.

In order to provide enough energy to the vacuum field, one may run inside the pool at enough speed. In this way the kinetic energy of the {\em particle} interacting with the Higgs field can be gained by such a field and few balls can scatter high enough to become visible out of the walls of the pool. Soon after the balls return to their original position, however the bottom line will be that people observing a certain region of the Universe in which particles are injected with enough energy may lead to the production of new particles, otherwise unobservable: these new particles can exist only for a very limited amount of time and are what physicists call the Higgs bosons.

\subsection{Caveats}
It is important that a public installation conveys the right messages and does not become merely a toy to play with. To this purpose, the shape and colour of the pool can be adjusted such that the attention of the visitors is maintained on the installation topic. For example, if the pool is narrow and long, such that visitors can only walk along its length, part of the playful aspect vanishes or, at least, diminishes. A similar effect can be obtained using monochromatic balls (and possibly of a neutral colour), in contrast with those found in playgrounds and shown in Fig.~\ref{fig:ballpool}.

On the other hand, the requirements in terms of size and shape of the model are not so stringent. Even if, as stated above, a narrow, long enough pool helps in conveying the right message, the proposed metaphor can be realised at almost any size. A tabletop setup, for example, can be used almost in the same way: in this case one can use a cart moved by hands to simulate the effects of a person walking in the pool. The   perception would be similar and understanding why the interaction lead to the appearance of a mass would be equally convincing.

\section{Summary}
We described a possible installation that can be used in scientific exhibits as well as in classrooms  to explain the physics of the Higgs boson to the grand public. Such an installation is easy to realise and cost effective and potentially addresses all the features of the Higgs field interaction with other particles. It is aimed at all ages and versatile enough to modulate the level of the analogy according to the audience background. 

\ack
I am indebted with the Director of Exhibits and Media Studio at the S. Francisco's Exploratorium Thomas Rockwell for his precious suggestions and his advices.

\end{document}